\documentclass[intlimits,twoside,a4paper]{article}

\usepackage{amsmath,amssymb}
\usepackage{graphicx}
\usepackage[T2A]{fontenc}
\usepackage[cp1251]{inputenc}

\usepackage[eqsecnum]{cmpj2}

\issue{2014}{17}{3}{33005}
\doinumber{10.5488/CMP.17.33005}

\newcommand{\ba}{\bar{a}}
\newcommand{\bb}{\bar{b}}

\title[Zubarev NSO method in Renyi
statistics. Reaction-diffusion processes]%
{Zubarev nonequilibrium statistical operator method in Renyi
statistics. Reaction-diffusion processes%
\thanks{In memory of Prof. A.~Olemskoi.}
}
\author[P.~Kostrobij \textsl{et al.}]%
{P.~Kostrobij\refaddr{label1}, R.~Tokarchuk\refaddr{label1}, M.~Tokarchuk\refaddr{label1,label2}, B.~Markiv\refaddr{label2}}
\addresses{
\addr{label1} National University ``Lviv Polytechnic'', 12~Bandera St.,
79013 Lviv, Ukraine
\addr{label2} Institute for Condensed Matter Physics
of the National Academy of Sciences of Ukraine, \\
1~Svientsitskii St., 79011 Lviv, Ukraine
}

\date{Received April 30, 2014, in final form May 29, 2014}
\authorcopyright{P.~Kostrobij, R.~Tokarchuk, M.~Tokarchuk, B.~Markiv, 2014}

\begin{document}
\maketitle

\begin{abstract}
The Zubarev nonequilibrium statistical operator (NSO) method in Renyi statistics is discussed.
The solution of $q$-parametrized Liouville equation within the NSO method is obtained.
A statistical approach for a consistent description
of reaction-diffusion processes in  ``gas-adsorbate-metal'' system
is proposed using the NSO method in Renyi statistics.
\keywords Renyi entropy, nonequilibrium statistical operator, chemical reaction
\pacs 05.45.-a, 82.20.-w
\end{abstract}

\section{Introduction}

Nowadays, investigating complex,
self-organizing, fractal structures and various physical phenomena,
such as subdiffusion, turbulence and chemical reactions, as well
as various economical, social and biological systems, Tsallis {\cite{1}}, Renyi {\cite{20,21}},
Sharma-Mittal \cite{34,35} statistics as well as superstatistics {\cite{36,37}}
are extensively used along with the Gibbs one.
A significant contribution to these investigations was made by A.I.~Olemskoi \cite{41a,42a,43a,44a,45a}
whose scientific activity we really lack.
In particular, problems in synergetic description of self-organizing systems,
description of the dynamics of phase transitions within the synergetic approach, the theory
of stochastic systems with singular multiplicative noise are elegantly presented in the original
work \cite{42a}.

The Tsallis entropy is widely used
in various directions of nonextensive statistical mechanics (for example, see  \cite{6,11,13} and references therein). Some
examples are the phenomena of subdiffusion {\cite{42,45}} and
turbulence \cite{47,48}, and the investigations
of transport coefficients in gases and plasma \cite{54}, as
well as quantum dissipative systems \cite{56}.
The energy fluctuations \cite{57}, kinetics of nonequilibrium plasma \cite{62},
problems of self-gravitating systems \cite{63}
and complex systems \cite{66,67}
were investigated within the Tsallis formalism. In references \cite{68,69,70}, Tsallis statistics was applied to
a description of chemical reactions, in particular, nonlinear equations of reaction-diffusion processes were obtained in reference \cite{68}. Despite the wide application of Tsallis entropy as a
generalization of Gibbs-Shannon entropy, the Renyi entropy is of
great interest as well {\cite{24,27,29,30,33,42,71,72,73}}. In particular, in this
case it is possible to determine a connection between the parameter $q$
and the heat capacity of the system \cite{29}. It is important to
note the papers \cite{74,75} by Luzzi et al.,
where the nonequilibrium statistical operator  method
and the Renyi entropy are used in describing the systems far
from equilibrium. In particular, the nonequilibrium
$q$-dependent Renyi ensemble as well as the generalized distribution
functions of bosons and fermions were obtained in reference \cite{74}. Therein the
experiments on anomalous luminescence at nanometer quantum dots in
semiconductor heterostructures were also described in this approach. A statistical
approach for a description of fractal physical-chemical systems
based on non-Fickian diffusion processes was proposed in reference \cite{75}.
Therein the investigations of anomalous diffusion in fractal-like electrodes in microbatteries were carried out.
{The nonextensive approach \cite{78a} as well as other ones \cite{79a,80a}
leading to Lindblad equation was used to describe a decoherence in quantum mechanics.}
The references \cite{79,80,81} are devoted to the investigation of nonlinear kinetics based on the Kramers,
Boltzmann and Fokker-Planck equations within the framework of generalized statistics.

In the present paper, an approach to the formulation of extensive
statistical mechanics of nonequilibrium processes~\cite{33} is considered, based on the Zubarev NSO method \cite{82,83,84}
and the maximum entropy principle for the Renyi entropy. This
statistical approach is applied to a consistent description of
reaction-diffusion processes in the ``gas-adsorbate-metal'' system.
Reaction-diffusion and adsorption-desorption processes on the
metal surface are nonlinear. They manifest an oscillation character,
possess memory effects and are actual in terms of nanostructure
formation on the surfaces occurring in catalytic phenomena \cite{85,86,87,88,89,90,91,92}.

\section{Renyi entropy and nonequilibrium statistical operator method}

The nonequilibrium state of a classical or quantum system
of interacting particles is completely described by the
nonequilibrium statistical operator $\varrho(x^N;t)$, which
satisfies the classical or quantum Liouville (von Neumann) equation:
\begin{eqnarray}
\label{math/2} \frac{\partial}{\partial t}\varrho(x^N;t)
+\ri L_N\varrho(x^N;t)=0.
\end{eqnarray}
Here, $\ri L_N$ is the Liouville operator of a system.

Within the NSO method framework, we will be looking for solutions
of equation (\ref{math/2}) which are independent of the initial
conditions. The solutions will explicitly depend  only on the
observable quantities
\begin{equation}
\int \rd\Gamma_N\hat{P}_n\varrho(x^N;t)=\langle\hat{P}_n\rangle^t.
\end{equation}
The nonequilibrium statistical operator has the following form:
\begin{eqnarray}
\label{math/26}\varrho(x^N;t)=\varrho_\textrm{rel}(x^N;t)-\int_{-\infty}^t
\re^{\varepsilon(t'-t)}T(t,t')
\left[1-P_\textrm{rel}(t')\right]\ri L_N\varrho_\textrm{rel}(x^N;t')\rd t',
\end{eqnarray}
where
\[
T(t,t')=\exp_+\left\{-\int_{t'}^t\left[1-P_\textrm{rel}(t')\right]\ri L_N\rd t'\right\}
\]
is the evolution operator containing the projection ($\exp_+$ denotes ordered exponential). The relevant
statistical operator (distribution function)
$\varrho_\textrm{rel}(x^N;t)$ will be determined using the maximum
entropy principle for the Renyi entropy
\begin{equation}
\label{math/3'} S^\textrm{R}(\varrho)=\frac{1}{1-q}\ln\int
\rd\Gamma_N\varrho^q(t)
\end{equation}
at fixed parameters of a reduced description, taking into
account the normalization condition.

The relevant statistical operator corresponding to the Renyi
entropy maximum has the following form:
\begin{align}
\label{math/14}
\varrho_\textrm{rel}(t)&=\frac{1}{Z_\textrm{R}}
\left[1-\frac{q-1}{q}\sum_nF_n(t)\delta\hat{P}_n\right]
^{\frac{1}{q-1}}, \\
%
\label{math/15}
Z_\textrm{R}(t)&=\int
\rd\Gamma_N\left[1-\frac{q-1}{q}\sum_nF_n(t)\delta\hat{P}_n\right]
^{\frac{1}{q-1}}.
\end{align}
$Z_\textrm{R}(t)$ is the partition function of the relevant statistical
operator,
$\delta\hat{P}_{n}=\hat{P}_{n}-\langle\hat{P}_n\rangle^t$. The
Lagrange multipliers $F_n(t)$ are defined from the
self-consistency conditions:
\begin{equation}
\label{math/15'}
\langle\hat{P}_n\rangle^t=\langle\hat{P}_n\rangle_\textrm{rel}^t\,,
\end{equation}
where $\langle\ldots\rangle^t_\textrm{rel}=\int
\rd\Gamma_N\ldots\varrho_\textrm{rel}(x^N;t)$.

In order to determine the generalized Kawasaki-Gunton projection operator entering (\ref{math/26})
\begin{eqnarray}
\label{math/18}
P_\textrm{rel}(t)\varrho'=\left(\varrho_\textrm{rel}(t)-\sum_n\frac{\delta
\varrho_\textrm{rel}(t)}{\delta
\langle\hat{P}_n\rangle^t}\langle\hat{P}_n\rangle^t\right)\int
d\Gamma_N \varrho'
+\sum_n\frac{\delta \varrho_\textrm{rel}(t)}{\delta
\langle\hat{P}_n\rangle^t}\int d\Gamma_N\hat{P}_n\varrho'
\end{eqnarray}
%
it is convenient to present the relevant statistical operator in a slightly
different form:
\begin{equation}
\label{math/14*}
\varrho^*_\textrm{rel}(t)=\frac{1}{Z^*_\textrm{R}}
\left[1-\frac{q-1}{q}\sum_nF^*_n(t)\hat{P}_n\right]
^{\frac{1}{q-1}},
\end{equation}
with the partition function
\begin{equation}
\label{math/15*} Z^*_\textrm{R}(t)=\int
\rd\Gamma_N\left[1-\frac{q-1}{q}\sum_nF^*_n(t)\hat{P}_n\right]
^{\frac{1}{q-1}}
\end{equation}
and
\begin{equation}
F^*_n(t)=\frac{F_n(t)}{1+\frac{q-1}{q}\sum_lF_l(t)\langle\hat{P}_l\rangle^t}\,.
\end{equation}
An action of the operators $P_\textrm{rel}(t)\ri L_N$ on the relevant
statistical operator can be presented as follows, by means of
generalized projection which  now acts on the dynamic variables,
$
P_\textrm{rel}(t)\ri L_N\varrho_\textrm{rel}(t)=P_\textrm{rel}(t)A(t)\varrho_\textrm{rel}(t)
=[P(t)A(t)]\varrho_\textrm{rel}(t),
$
where
\begin{eqnarray}
\label{math/39} P(t)\ldots=\sum_{m,n}\langle\ldots\hat{P}_m\rangle_\textrm{rel}
\langle\hat{P}_m\delta\{[q\psi(t)]^{-1}\hat{P}_n\}\rangle_\textrm{rel}^{-1}
\delta\{[q\psi(t)]^{-1}\hat{P}_n\},
\end{eqnarray}
\begin{eqnarray*}
\label{math/39'} \psi(t)=1-\frac{q-1}{q}\sum_nF^*_n(t)\hat{P}_n\,.
\end{eqnarray*}
Taking into account that
$[1-P_\textrm{rel}(t)]\ri L_N\varrho_\textrm{rel}(t)=-\sum_nI_n(t)F_n(t)\varrho_\textrm{rel}(t)$,
where
\begin{eqnarray}
\label{math/43} I_n(t)=[1-P(t)]
\frac{1}{q}\psi^{-1}(t)\dot{\hat{P}}_{n}
\end{eqnarray}
are the generalized flows, we can now write down an explicit
expression for the \emph{nonequilibrium statistical operator}
\begin{eqnarray}
\label{math/269}
\varrho(x^N;t)=\varrho_\textrm{rel}(x^N;t)+\sum_n\int_{-\infty}^t\re^{\varepsilon(t'-t)}T(t,t')
 I_n(t')F_n(t')\varrho_\textrm{rel}(x^N;t')\rd t'.
\end{eqnarray}
This allows us to obtain generalized \emph{transport
equations} for the reduced-description parameters. They can be
presented in the form
\begin{eqnarray}
\label{math/44}\frac{\partial}{\partial
t}\langle\hat{P}_m\rangle^t=\langle\dot{\hat{P}}_m\rangle^t_\textrm{rel}
+\sum_n\int_{-\infty}^t\re^{\varepsilon(t'-t)}\varphi_{mn}(t,t')F_n(t')\rd t',
\end{eqnarray}
with the generalized \emph{transport kernels} (memory functions)
\begin{eqnarray}
\label{math/45}\varphi_{mn}(t,t')= \int
\rd\Gamma_N\left\{\dot{\hat{P}}_mT(t,t')I_n(t')\varrho_\textrm{rel}(t')\right\}
\end{eqnarray}
which describe the dissipative processes in the system.

\section{$q$-generalization of Liouville equation}

An interesting generalization of Liouville equation was proposed in \cite{78a}, where the $q$-parametrized Liouville equation was obtained:
\begin{eqnarray}
\label{math/2a} \frac{\partial}{\partial t}\varrho(x^N;t)
+\ri\tilde{L}_N(t)\varrho(x^N;t)=0,
\end{eqnarray}
where
\begin{eqnarray}
\label{math/3a}
\ri\tilde{L}_N(t)= \frac{\ri L_N}{1+(1-q)t\ri L_N}
\end{eqnarray}
is the $q$-parametrized Liouville operator. When $q=1$ we have $\ri\tilde{L}_N(t)=\ri L_N$. For $|1-q|\Omega t\ll 1$, where $\Omega$ is the
characteristic frequency of the physical system considered, we may write~\cite{78a} for (\ref{math/2a}):
\begin{eqnarray}
\label{math/2aa} \frac{\partial}{\partial t}\varrho(x^N;t)
+\left[\ri L_N-(1-q)t(\ri L_{N})^{2}\right]\varrho(x^N;t)=0.
\end{eqnarray}
This is the Lindblad type equation for nonequilibrium statistical operator $\varrho(x^N;t)$. The Lindblad type equation within Renyi statistics was obtained in \cite{56}.

The solution of $q$-parametrized Liouville equation within the NSO method can be presented as follows:
\begin{eqnarray}
\label{math/4a}
\varrho(x^N;t)&=&-\varepsilon \int_{-\infty}^t \re^{\varepsilon(t'-t)}T_{q}(t,t')
\varrho_\textrm{rel}(x^N;t')\rd t'\nonumber\\
&=&\varrho_\textrm{rel}(x^N;t)-\int_{-\infty}^t\re^{\varepsilon(t'-t)}\left\{ \vphantom{\int_{0}^{1}}T_{q}(t,t')\frac{\partial }{\partial t'}\right.\nonumber\\
&&-
\left.\int_{0}^{1}T_{q}^{\tau}(t,t')\frac{\ri L_N}{1+(1-q)(t'-t)\ri L_N}T_{q}^{1-\tau}(t,t')d\tau\right\}
\varrho_\textrm{rel}(x^N;t')\rd t',
\end{eqnarray}
where
\begin{eqnarray}
\label{math/5a}
T_{q}(t,t')= \exp_+\left\{-\int_{t'}^{t}\frac{\ri L_N}{1+(1-q)t''\ri L_N}\rd t''\right\}
\end{eqnarray}
is the parametrized evolution operator. For $|1-q|\Omega t\ll 1$ from (\ref{math/4a}) we obtain the solution of Lindblad type equation for  $\varrho(x^N;t)$.

It is important to note that, at $q\rightarrow t$, from (\ref{math/269}) and (\ref{math/44})
we reproduce the nonequilibrium statistical operator and the generalized transport equations
for the reduced-description parameters within Gibbs statistics \cite{82,83,84}.
In the following section we apply the discussed approach to a description of reaction-diffusion
processes, in particular, in catalytic processes.

\section{Reaction-diffusion processes}

Let us start with the Hamiltonian of the system of
``gas-adsorbate-metal'' in the following form $H=H'+H_\textrm{reac}$,
$H'=H_{a}+H_{a}^\textrm{int}$. Here, $H_a$ is the Hamiltonian of the gas
subsystem considered according to the classical approach; $H_{a}^\textrm{int}$
is the Hamiltonian describing the interaction between the gas atoms and
the atoms adsorbed on the metal surface; $H_\textrm{reac}$ is the Hamiltonian
of interaction for chemical reactions between atoms or molecules
adsorbed on the metal surface \cite{85}
\begin{eqnarray}
H_\textrm{reac}=\sum_{\ba,\bb,\ba',\bb'}
\left(\langle\ba',\bb'|\Phi_\textrm{reac}|\ba,\bb\rangle
\hat{q}_{\ba'}^{+}\hat{q}_{\bb'}^{+}\hat{q}_{\ba}\hat{q}_{\bb}
+\langle\ba',\bb'|\Phi_\textrm{reac}|\ba,\bb\rangle^*
\hat{q}_{\ba}^{+}\hat{q}_{\bb}^{+}\hat{q}_{\ba'}\hat{q}_{\bb'}
\right),
\end{eqnarray}
where $\langle\ba',\bb'|\Phi_\textrm{reac}|\ba,\bb\rangle
=\langle\ba,\bb|\Phi_\textrm{reac}|\ba',\bb'\rangle$ are the amplitudes
of reaction between reagents $A$ and $B$ (supposed to be known
from quantum mechanics). We introduce the notation $\ba,\bb$ and
$\ba',\bb'$ for state of the reagents $A$, $B$ (atoms or molecules)
and for the state of atoms in the reaction products $AB$. Here,
$\hat{q}_{\ba'}^{+}$, $\hat{q}_{\bb'}^{+}$, $\hat{q}_{\ba}^{+}$,
$\hat{q}_{\bb}^{+}$ and $\hat{q}_{\ba'}$, $\hat{q}_{\bb'}$,
$\hat{q}_{\ba}$, $\hat{q}_{\bb}$ are the operators of creation and
annihilation of atomic states  $\ba',\bb'$ for molecule $AB$,
 and $\ba,\bb$ for $A$ and $B$, respectively.

Parameters of the reduced description are the averaged densities of
gas atoms absorbed and not absorbed on the metal surface
$\langle\hat{n}_{\ba}(\vec{R})\rangle^t=\textrm{Sp}\left\{\hat{n}_{\ba}(\vec{R})\varrho(t)\right\}$,
$\langle\hat{n}_{a}(\vec{R})\rangle^t=\textrm{Sp}\left\{\hat{n}_{a}(\vec{R})\varrho(t)\right\}$.
$\hat{n}_{\ba}(\vec{R})$ is the density operator for gas atoms
adsorbed on the metal surface in the state $\nu$;
$
\hat{n}_{\ba}(\vec{R})=\sum_{j}^{N_a^\textrm{ad}}\hat{\psi}_{\nu
j}^+(\vec{R})\hat{\psi}_{\nu j}(\vec{R})$;
$\hat{\psi}_{\nu
j}^+(\vec{R})$, $\hat{\psi}_{\nu j}(\vec{R})$ are the creation and
annihilation operators of gas atoms absorbed in the state $\nu$ on the
metal surface which satisfy the Bose-type commutation relations.
Since we do not consider a catalyst surface explicitly in this
model, the states $\nu$ and $\mu$ mean the adsorption centers,
where atoms can be located.
$\hat{n}_{a}(\vec{r})=\sum_{j}^{N_a^\textrm{ad}}\delta(\vec{r}-\vec{r}_j)$
 is the microscopic density of gas atoms,
$\langle\hat{G}_{\ba\bb}^{\nu\mu}(\vec{R},\vec{R}')\rangle^t
=\textrm{Sp}\big\{\hat{G}_{\ba\bb}^{\nu\mu}(\vec{R},\vec{R}')\varrho(t)\big\}$
 is the nonequilibrium pair distribution function of the atoms or
molecules adsorbed on the metal surface, and
$\hat{G}_{\ba\bb}^{\nu\mu}(\vec{R},\vec{R}')
=\hat{n}_{\ba}^{\nu}(\vec{R})\hat{n}_{\bb}^{\mu}(\vec{R}')$.

The relevant statistical operator has the form:
\begin{eqnarray}\lefteqn{
\varrho_\textrm{rel}=\frac{1}{Z_\textrm{R}(t)}\left\{1-\frac{q-1}{q}\beta\left[\vphantom{\sum_{\ba\bb}}\delta
H(t)-\sum_a\int \rd\vec{r}\mu_a(\vec{r};t)\delta
\hat{n}_a(\vec{r};t) \right.\right. }\nonumber
\\&&\mbox{}\left.\left.-\sum_{\ba}\sum_{\nu}\int
\rd\vec{R}\mu_{\ba}^{\nu}(\vec{R};t)\delta
\hat{n}_{\ba}^{\nu}(\vec{R};t)-\sum_{\ba\bb}\sum_{\nu\mu}\int
\rd\vec{R}\rd\vec{R}'M_{\ba\bb}^{\nu\mu}(\vec{R},\vec{R}';t)
\hat{G}_{\ba\bb}^{\nu\mu}(\vec{R},\vec{R}';t)\right]\right\}^{\frac{1}{q-1}}.
\end{eqnarray}

The partition function of the relevant statistical operator is as follows
\begin{eqnarray}\lefteqn{
Z_\textrm{R}(t)=\textrm{Sp}\left\{1-\frac{q-1}{q}\beta\left[\vphantom{\sum_{\ba\bb}}\delta H(t)-\sum_a\int
\rd\vec{r}\mu_a(\vec{r};t)\delta \hat{n}_a(\vec{r};t) \right.\right.
}\nonumber
\\&&\mbox{}\left.\left.-\sum_{\ba}\sum_{\nu}\int
\rd\vec{R}\mu_{\ba}^{\nu}(\vec{R};t)\delta
\hat{n}_{\ba}^{\nu}(\vec{R};t)-\sum_{\ba\bb}\sum_{\nu\mu}\int
\rd\vec{R}\rd\vec{R}'M_{\ba\bb}^{\nu\mu}(\vec{R},\vec{R}';t)
\hat{G}_{\ba\bb}^{\nu\mu}(\vec{R},\vec{R}';t)\right]\right\}^{\frac{1}{q-1}}.
\end{eqnarray}
Here,
\[\textrm{Sp}\{\ldots\}=\prod_{\alpha}
\int\frac{(\rd\vec{r}\rd\vec{p})^{N_{\alpha}}}{N_{\alpha}!(2\pi\hbar)^{3N_{\alpha}}}
\textrm{Sp}_{(\nu,\xi,\sigma)}\{\ldots\},\]
$N_{\alpha}=\{N_a,N_{\ba}\}$, $\textrm{Sp}_{(\nu,\xi,\sigma)}$ means the averaged summation over all values of spin and quantum numbers.
The parameters $\mu_a(\vec{r};t)$, $\mu_{\ba}^{\nu}(\vec{R};t)$,
$M_{\ba\bb}^{\nu\mu}(\vec{R},\vec{R}';t)$ should be determined
from the corresponding self-consistency conditions
\begin{eqnarray*}
\langle\hat{n}_{a}(\vec{r})\rangle^t=\langle\hat{n}_{a}(\vec{r})\rangle^t_\textrm{rel}\,,
\qquad %
\langle\hat{n}_{\ba}^{\nu}(\vec{R})\rangle^t
=\langle\hat{n}_{\ba}^{\nu}(\vec{R})\rangle^t_\textrm{rel}\,,
\qquad %
\langle\hat{G}_{\ba\bb}^{\nu\mu}(\vec{R},\vec{R}')\rangle^t
=\langle\hat{G}_{\ba\bb}^{\nu\mu}(\vec{R},\vec{R}')\rangle^t_\textrm{rel}\,,
\end{eqnarray*}
hence, we found that $\mu_a(\vec{r};t)$ defines a local
chemical potential of gas atoms; $\mu_{\ba}^{\nu}(\vec{R};t)$ is a
local chemical potential of an atom adsorbed in a state $\nu$ on
the metal surface. $\delta H(t)=H-\langle H\rangle^t$, $\delta
\hat{n}_{a}(\vec{r};t)=\hat{n}_{a}(\vec{r})-\langle\hat{n}_{a}(\vec{r})\rangle^t$,
$\delta\hat{n}_{\ba}^{\nu}(\vec{R};t)=\hat{n}_{\ba}^{\nu}(\vec{R})
-\langle\hat{n}_{\ba}^{\nu}(\vec{R})\rangle^t$,
$\delta\hat{G}_{\ba\bb}^{\nu\mu}(\vec{R},\vec{R}';t)
=\hat{G}_{\ba\bb}^{\nu\mu}(\vec{R},\vec{R}')
-\langle\hat{G}_{\ba\bb}^{\nu\mu}(\vec{R},\vec{R}')\rangle^t$.

According to (\ref{math/269}), NSO of the system
``gas-adsorbate-metal'' has the following form
\begin{eqnarray}
\label{math/3.4}
\lefteqn{\varrho(t)=\varrho(t)_\textrm{rel}+\sum_a\int
\rd\vec{r}\int_{-\infty}^t\re^{\varepsilon(t'-t)}T(t,t')\left[\int_0^1
\rd\tau\varrho_\textrm{rel}^{\tau}(t')I_a(\vec{r};t)\varrho_\textrm{rel}^{1-\tau}(t')\beta\mu_a(\vec{r};t')
\right]\rd t'}\nonumber
\\&&\mbox{}+\sum_{\ba}\sum_{\nu}\int
\rd\vec{R}\int_{-\infty}^t\re^{\varepsilon(t'-t)}T(t,t')\left[\int_0^1
\rd\tau\varrho_\textrm{rel}^{\tau}(t')I_{\ba}^{\nu}(\vec{R};t)\varrho_\textrm{rel}^{1-\tau}(t')
\beta\mu_{\ba}^{\nu}(\vec{R};t')\right]\rd t'\nonumber
\\&&\mbox{}+\sum_{\ba,\bb}\sum_{\nu,\mu}\int
\rd\vec{R}\rd\vec{R}'\int_{-\infty}^t\re^{\varepsilon(t'-t)}T(t,t')\left[\int_0^1
\rd\tau\varrho_\textrm{rel}^{\tau}(t')I_{G_{\ba\bb}}^{\nu\mu}(\vec{R},\vec{R}';t)
\varrho_\textrm{rel}^{1-\tau}(t')\beta
M_{\ba\bb}^{\nu\mu}(\vec{R},\vec{R}';t')\right]\rd t'.
\end{eqnarray}
\begin{eqnarray}\label{math/3.5}
I_a(\vec{r};t)=[1-P(t)]\frac{1}{q}\psi^{-1}(t)\dot{\hat{n}}_a(\vec{r}),\qquad
I_{\ba}^{\nu}(\vec{R};t)=[1-P(t)]\frac{1}{q}\psi^{-1}(t)\dot{\hat{n}}_{\ba}^{\nu}(\vec{R}),
\nonumber \\
I_{G_{\ba\bb}}^{\nu\mu}(\vec{R},\vec{R}';t)
=[1-P(t)]\frac{1}{q}\psi^{-1}(t)\dot{\hat{G}}_{\ba\bb}^{\nu\mu}(\vec{R},\vec{R}')\qquad\qquad\qquad
\end{eqnarray}
are the generalized flows describing reaction-diffusion processes.
The function $\psi(t)$ is defined by the relation
\begin{eqnarray}\lefteqn{
\psi(t)=1-\frac{q-1}{q}\beta\left[\vphantom{\sum_{\ba\bb}}\delta H(t)-\sum_a\int
\rd\vec{r}\mu_a(\vec{r};t)\delta \hat{n}_a(\vec{r};t) \right. }\nonumber
\\&&\mbox{}\left.-\sum_{\ba}\sum_{\nu}\int
\rd\vec{R}\mu_{\ba}^{\nu}(\vec{R};t)\delta
\hat{n}_{\ba}^{\nu}(\vec{R};t)-\sum_{\ba\bb}\sum_{\nu\mu}\int
\rd\vec{R}\rd\vec{R}'M_{\ba\bb}^{\nu\mu}(\vec{R},\vec{R}';t)
\hat{G}_{\ba\bb}^{\nu\mu}(\vec{R},\vec{R}';t)\right].
\end{eqnarray}
By means of NSO (\ref{math/3.4}), we obtain the set of self-consistent
generalized transport equations for averaged densities of adsorbed
and non-absorbed atoms along with nonequilibrium pair distribution
function of absorbed atoms (or molecules). It has the following
form
\begin{eqnarray}\label{math/3.7}\lefteqn{
\frac{\partial}{\partial t}
\langle\hat{n}_{a}(\vec{r})\rangle^t=\langle\dot{\hat{n}}_{a}(\vec{r})\rangle^t_\textrm{rel}
+\sum_b\int
\rd\vec{r}'\int_{-\infty}^t\re^{\varepsilon(t'-t)}\varphi_{n_an_b}(\vec{r},\vec{r}';t,t')
\beta\mu_{b}(\vec{r}';t')\rd t'}\nonumber
%
%
\\&&\mbox{}+\sum_{\bb}\sum_{\nu'}\int
\rd\vec{R}'\int_{-\infty}^t\re^{\varepsilon(t'-t)}
\varphi_{n_an_{\bb}}^{\nu'}(\vec{r},\vec{R}';t,t')
\beta\mu_{\bb}^{\nu'}(\vec{R}';t')\rd t'    \nonumber
\\&&\mbox{}+\sum_{\ba'\bb}\sum_{\nu'\mu'}\int
\rd\vec{R}'\rd\vec{R}''\int_{-\infty}^t\re^{\varepsilon(t'-t)}
\varphi_{n_aG_{\ba'\bb}}^{\nu'\mu'}(\vec{r},\vec{R}',\vec{R}'';t,t')
\beta M_{\ba'\bb}^{\nu'\mu'}(\vec{R}',\vec{R}'';t')\rd t',
\end{eqnarray}
%
%
\begin{eqnarray}\label{math/3.8}\lefteqn{
\frac{\partial}{\partial t}
\langle\hat{n}_{\ba}^{\nu}(\vec{R})\rangle^t
=\langle\frac{1}{\ri\hbar}[{\hat{n}}_{\ba}^{\nu}(\vec{R}),H']\rangle^t_\textrm{rel}
+\langle\frac{1}{\ri\hbar}[{\hat{n}}_{\ba}^{\nu}(\vec{R}),H_\textrm{reac}]\rangle^t_\textrm{rel}}
\nonumber
\\&&\mbox{}+\sum_b\int
\rd\vec{r}'\int_{-\infty}^t\re^{\varepsilon(t'-t)}
\varphi_{n_{\ba}n_b}^{\nu}(\vec{R},\vec{r}';t,t')
\beta\mu_{b}(\vec{r}';t')\rd t' \nonumber
\\&&\mbox{}+\sum_{\bb}\sum_{\nu'}\int
\rd\vec{R}'\int_{-\infty}^t\re^{\varepsilon(t'-t)}
\varphi_{n_{\ba}n_{\bb}}^{\nu\nu'}(\vec{R},\vec{R}';t,t')
\beta\mu_{\bb}^{\nu'}(\vec{R}';t')\rd t'    \nonumber
\\&&\mbox{}+\sum_{\ba'\bb}\sum_{\nu'\mu'}\int
\rd\vec{R}'\rd\vec{R}''\int_{-\infty}^t\re^{\varepsilon(t'-t)}
\varphi_{n_{\ba}G_{\ba'\bb}}^{\nu'\mu'}(\vec{r},\vec{R}',\vec{R}'';t,t')
\beta M_{\ba'\bb}^{\nu'\mu'}(\vec{R}',\vec{R}'';t')\rd t',
\end{eqnarray}
%
%
\begin{eqnarray}\label{math/3.9}\lefteqn{
\frac{\partial}{\partial t}
\langle\hat{G}_{\ba\bb}^{\nu\mu}(\vec{R},\vec{R}')\rangle^t
=\langle\frac{1}{\ri\hbar}[\hat{G}_{\ba\bb}^{\nu\mu}(\vec{R},\vec{R}'),H']\rangle^t_\textrm{rel}
+\langle\frac{1}{\ri\hbar}[\hat{G}_{\ba\bb}^{\nu\mu}(\vec{R},\vec{R}'),H_\textrm{reac}]\rangle^t_\textrm{rel}}
\nonumber
\\&&\mbox{}+\sum_{b'}\int
\rd\vec{r}'\int_{-\infty}^t\re^{\varepsilon(t'-t)}
\varphi_{G_{\ba\bb}n_{b'}}^{\nu\mu}(\vec{R},\vec{R}';t,t')
\beta\mu_{b'}(\vec{r}';t')\rd t' \nonumber
\\&&\mbox{}+\sum_{\bb}\sum_{\nu'}\int
\rd\vec{R}''\int_{-\infty}^t\re^{\varepsilon(t'-t)}
\varphi_{G_{\ba\bb}n_{\bb'}}^{\nu\mu\nu'}(\vec{R},\vec{R}',\vec{R}'';t,t')
\beta\mu_{\bb}^{\nu'}(\vec{R}';t')\rd t'    \nonumber
\\&&\mbox{}+\sum_{\ba'\bb'}\sum_{\nu'\mu'}\int
\rd\vec{R}''\rd\vec{R}'''\int_{-\infty}^t\re^{\varepsilon(t'-t)}
\varphi_{G_{\ba\bb}G_{\ba'\bb}}^{\nu\mu\nu'\mu'}(\vec{r},\vec{R}',\vec{R}'',\vec{R}''';t,t')
\beta M_{\ba'\bb'}^{\nu'\mu'}(\vec{R}'',\vec{R}''';t')\rd t',
\end{eqnarray}
where  $\varphi_{n_{a}n_{b}}$,
$\varphi_{n_{\ba}n_{\bb}}^{\nu\nu'}$,
$\varphi_{n_{a}n_{\bb}}^{\nu}$,
$\varphi_{G_{\ba\bb}G_{\ba'\bb}}^{\nu\mu\nu'\mu'}$ are the
generalized transport kernels. The second term in the right-hand
side of~(\ref{math/3.8}),
$\langle\frac{1}{\ri\hbar}[{\hat{n}}_{\ba}^{\nu}(\vec{R}),H_\textrm{reac}]\rangle^t_\textrm{rel}$,
defines an averaged value of the operator of the rate of the reaction
between adsorbed atoms on the metal surface. Transport kernels are
built on the generalized flows~(\ref{math/3.5}) taking into account
the contributions of amplitudes of chemical reactions in the flows
$I_{\ba}^{\nu}(\vec{R};t')$ and
$I_{G_{\ba\bb}}^{\nu\mu}(\vec{R},\vec{R}';t')$, and have the
following form
\begin{eqnarray}
\varphi_{BB'}=\textrm{Sp}\left\{\dot{\hat{P}}_BT(t,t')\int_0^1
\rd\tau\varrho_\textrm{rel}^{\tau}(t')I_{B'}(t')\varrho_\textrm{rel}^{1-\tau}(t')\right\}.
\end{eqnarray}
In particular, $\varphi_{n_{a}n_{b}}(\vec{r},\vec{r}';t,t')$
describes dynamical correlations of diffusive flows of gas atoms
and is connected to the inhomogeneous diffusion coefficient of
atoms (or molecules) $D_{ab}(\vec{r},\vec{r}';t)$. Similarly, the
transport kernel
$\varphi_{n_{\ba}n_{\bb}}^{\nu\nu'}(\vec{R},\vec{R}';t,t')$
describes dynamical dissipative correlations of diffusive flows of
atoms adsorbed in the states $\nu$ and $\nu'$ on the metal surface and
determines an inhomogeneous diffusion coefficient of the atoms adsorbed
on the metal surface $D_{\ba\bb}^{\nu\nu'}(\vec{R},\vec{R}';t)$.
Transport kernel
$\varphi_{n_{\ba}n_{b}}^{\nu}(\vec{R},\vec{r}';t,t')$,
$\varphi_{n_{a}n_{\bb}}^{\nu'}(\vec{r},\vec{R}';t,t')$ describes
dynamical dissipative correlations between the flows of gas atoms and
the atoms adsorbed on the metal surface, and determines the inhomogeneous
coefficient of mutual diffusion ``gas atom-adsorbed atom''
$D_{a\bb}^{\nu'}(\vec{r},\vec{R}';t)$. It is very important to investigate these
diffusion coefficients. The transport kernel
$\varphi_{G_{\ba\bb}p}^{\nu\mu}$ ($p=n,\bar{n}$) describes
dissipative correlations of the flows and densities of adsorbed atoms
with the flows of atoms, molecules and adsorbed atoms. Memory function
$\varphi_{G_{\ba\bb}G_{\ba'\bb}}^{\nu\mu\nu'\mu'}
(\vec{R},\vec{R}',\vec{R}'',\vec{R}''';t,t')$ describes
reaction-diffusion processes between the atoms adsorbed on the metal
surface. They are higher memory functions with respect to
dynamical variables $G_{\ba\bb}^{\nu\mu}$. We note here that at $q\rightarrow 1$,
the generalized equations of reaction-diffusion processes correspond to the ones within
the Gibbs statistics.

\vspace{-1mm}

\section{Conclusions}

Summarizing, we proposed an approach to the formulation of extensive statistical mechanics
of nonequilibrium processes based on the Zubarev NSO method and maximum entropy principle
for Renyi entropy. We consider a $q$-parametrized Liouville equation which leads to Lindblad
type equation at $|1-q|\Omega t\ll 1$. The solution of this $q$-parametrized Liouville
equation is obtained within the NSO method. The proposed approach is used to describe
the reaction-diffusion processes which are relevant in catalytic nanotechnologies.
Consequently, we obtain generalized transport equations~(\ref{math/3.7})--(\ref{math/3.9})
for the nonequilibrium averaged densities of adsorbed and non-adsorbed atoms of consistent description of the reaction-diffusion
processes in the system ``gas-adsorbate-metal'' within Renyi
statistics. At $q=1$, these equations coincide with the equations
of reaction-diffusion processes within Gibbs statistics~\cite{85}.
As we can see, these equations are nonlinear and spatially
inhomogeneous. They can describe strong as well as  weak
nonequilibrium processes in a system. The application of the obtained results
to the consideration of a particular physical model will be done in our forthcoming works.

\clearpage

\ukrainianpart

\title{Метод нерівноважного статистичного оператора Зубарєва у статистиці Рені. Реакційно-дифузійні процеси}
\author{П. Костробій\refaddr{label1}, Р.~Токарчук\refaddr{label1}, М.~Токарчук\refaddr{label1,label2}, Б.~Марків\refaddr{label2}}
\addresses{
\addr{label1} Національний університет ``Львівська політехніка'',
вул.~С.~Бандери,~12, 79013 Львів, Україна
\addr{label2} Інститут фізики конденсованих систем НАН України,
вул. І. Свєнціцького, 1, 79011 Львів, Україна
}

\makeukrtitle

\begin{abstract}
\tolerance=3000%
Обговорюється метод нерівноважного статистичного оператора (НСО) Зубарєва у статистиці Рені. Отримано розв'язок $q$-параметризваного рівняння Ліувілля в рамках методу НСО.
Запропоновано статистичний підхід до узгодженого опису реакційно-дифузійних процесів у системі ``газ-адсорбат-метал'' методом НСО у статистиці Рені.
\keywords ентропія Рені, нерівноважний статистичний оператор, хімічні реакції
\end{abstract}

\end{document}